\documentclass[twocolappendix]{emulateapj}

\usepackage[usenames,dvipsnames]{color}
\usepackage{comment}

\shorttitle{Jet-Ejecta Interaction} 
\shortauthors{Duffell et al.}

\begin{document}

\title{Jet Dynamics in Compact Object Mergers: GW170817 Likely had a Successful Jet}

\author{Paul C. Duffell, Eliot Quataert, Daniel Kasen, and Hannah Klion}
\affil{Astronomy Department and Theoretical Astrophysics Center, University of California, Berkeley}
\email{duffell@berkeley.edu}

\begin{abstract}

We use relativistic hydrodynamic numerical calculations to study the interaction between a jet and a homologous outflow produced dynamically during binary neutron star mergers.  We quantify how the thermal energy supplied by the jet to the ejecta and the ability of a jet to escape the homologous ejecta depend on the parameters of the jet engine and the ejecta.  
For collimated jets initiated at early times compared to the engine duration, we show that successful breakout of the forward cocoon shock necessitates a  jet that successfully escapes the ejecta.  This is because the ejecta is expanding and absorbing thermal energy, so that the forward shock from a failed jet stalls before it reaches the edge of the ejecta.  This conclusion can be circumvented only for very energetic wide angle jets, with parameters that are uncomfortable given short-duration GRB observations.
For successful jets, we find two regimes of jet breakout from the ejecta, early breakout on timescales shorter than the engine duration, and late breakout well after the engine shuts off.  A late breakout can explain the observed delay between gravitational waves and gamma rays in GW 170817.  We show that for the entire parameter space of jet parameters surveyed here (covering energies $\sim 10^{48}-10^{51}$ ergs and opening angles $\theta_j \sim 0.07-0.4$) the thermal energy deposited into the ejecta by the jet propagation is less than that produced by r-process heating on second timescales by a factor of $\gtrsim 10$.   Shock heating is thus energetically subdominant in setting the luminosity of thermally powered transients coincident with neutron star mergers (kilonovae).  For typical short GRB jet parameters, our conclusion is stronger:  there is little thermal energy in the cocoon,  much less than what is needed to explain the early blue component of the kilonova in GW 170817.

\end{abstract}


\section{Introduction} \label{sec:intro}

The first detection of merging neutron stars by gravitational waves \citep[GW 170817,][]{2017PhRvL.119p1101A} was accompanied by a wealth of electromagnetic counterparts.  Notable was the first unambiguous detection of kilonova emission, which demonstrated that a significant amount of mass ($\gtrsim 0.04 M_{\sun}$) had been ejected with a velocity of $\sim 0.2 c$ and kinetic energy of $\sim 10^{51}$ ergs \citep[e.g.][]{2017ApJ...848L..12A, 2017Sci...358.1556C, 2017Sci...358.1559K, 2017Sci...358.1574S, 2017ApJ...848L..16S, 2017ApJ...848L..24V}.  

1.7 seconds after the merger, a low-energy short-duration gamma ray burst (GRB) was also detected \citep{2017ApJ...848L..13A}.  Presently there are two leading theories concerning the origin of the gamma rays: either they are produced by shock breakout from some injected energy from a central engine \citep{2018MNRAS.475.2971B, 2018MNRAS.tmp.1404G} or by a traditional short GRB observed off-axis \citep{2017MNRAS.471.1652L, 2017arXiv171203237L, 2018MNRAS.473L.121K}.  Either of these models requires  relativistic material to escape through the $v \sim 0.2$c ejecta, at least some fraction of which is relatively spherically distributed \citep{2013PhRvD..87b4001H}.

The shock breakout scenario is sometimes envisioned as being associated with a failed GRB jet \citep[e.g.][]{2018MNRAS.tmp.1404G}.  In this scenario, despite the failure of the jet to escape the surrounding ejecta, because significant  energy was injected into the flow in the form of a cocoon, a shock continues to propagate forward, eventually breaking out of the ejecta, accelerating to relativistic velocities and releasing gamma rays.

The thermal energy imparted to the cocoon may also be important for the quasi-thermal kilonova emission in neutron star mergers.  For GW 170817 in particular, there are multiple observed components to the thermal transient \citep{2017Sci...358.1574S}; a blue component with faster velocities ($\sim 0.2c$), a red component with slower velocities ($\sim 0.1c$), and possibly an intermediate component, see \cite{2017ApJ...851L..21V}.  It has been suggested that the blue component (particularly at early times) may be partially or fully powered by thermal energy in the cocoon \citep{2017Sci...358.1559K, 2018ApJ...855..103P}, rather than by the radioactive decay of neutron-rich elements which is believed to power the majority of the thermal kilonova  emission \citep{2010MNRAS.406.2650M, 2013ApJ...775...18B}.

The jet-ejecta interaction in neutron star mergers is somewhat analogous to a similar scenario in long-duration GRBs; there, the GRB jet needs to punch its way through a collapsing star \citep{1999ApJ...524..262M, 2017arXiv170802630B}.  However, the scenario is not equivalent.  The dynamical ejecta from a neutron star merger is a homologously expanding outflow whose velocity cannot be neglected, as opposed to the collapsar whose infall velocities are slow enough that it can be effectively considered stationary.  This difference qualitatively affects the dynamics of the interaction.  In this work, we perform analytical and numerical calculations of the jet-ejecta interaction, to develop new scalings for the conditions under which jets can break out of the ejecta, the resulting breakout time of the jet, and the thermalized jet energy as a function of time. 

We find several important results.  First, assuming a narrow engine whose energy is subdominant relative to the total kinetic energy in the dynamical ejecta, and assuming the jet is initiated at early times compared with the breakout time, shock breakout of the cocoon only occurs if the jet is also successful.  There is no shock breakout for a failed jet with narrow injection angle.  This is because the thermal energy from a failed jet is absorbed into the dynamical ejecta; because the ejecta is expanding, a low-energy shock never makes it to the edge of the outflow.  Secondly, for a fixed total energy injection, the condition for success is independent of the duration of the engine; this is a consequence of scale invariance.  Therefore, success or failure of the jet only depends on the total energy injected and the solid angle within which it is injected.  Third, a consequence of this is that, for very energetic jets, the breakout time is inversely proportional to the jet power.  Fourth, we find two distinct regimes, ``early breakout" and ``late breakout".  ``Late breakout" jets can break out {\em after} the engine turns off.  Fifth, the amount of energy that is thermalized can be estimated straightforwardly from the above considerations, leading to a robust prediction for the thermal energy in the cocoon at late times.  For typical short GRB parameters, this thermal energy is very low.

Analytical scalings for this problem are derived in Section \ref{sec:anly}.  The numerical set-up is briefly described in Section \ref{sec:numerics}.  Results of numerical calculations and comparison with the analytical scalings are presented in Section \ref{sec:results}.  Finally, these results are discussed in the context of short GRBs and GW170817 observations in Section \ref{sec:disc}.

\section{Analytical Scalings} \label{sec:anly}

\subsection{Success vs. Failure}

There are multiple components to the ejecta in binary neutron star mergers, including tidal tails (e.g., \citealt{1999AnA...341..499R}), quasi-spherical ejecta produced when the neutron stars first collide and shock (e.g., \citealt{2013PhRvD..87b4001H}), and accretion disk winds (e.g., \citealt{2008MNRAS.390..781M}).   The jet is most likely to directly interact with the quasi-spherical dynamical ejecta.   This can be modeled as a homologously expanding medium, which is qualitatively distinct from the stationary medium typically considered in the context of active galactic nuclei or long GRB jets.  One notable distinction is that a homologous outflow has no physical length scale, whereas a long GRB progenitor has a fixed radius.  The GRB engine has only one timescale, the duration of the engine (assuming there is no significant delay between the ejection of the outflow and injection of the jet).  Therefore, any physical length or time scale in the solution is proportional to the engine duration; a calculation with the same energy injected but double the engine duration would yield an identical solution, but with all physical length and timescales multiplied by two.

The jet-ejecta interaction in neutron star mergers is thus less like the jet-collapsar interaction in long GRBs, and more like a pulsar-ejecta interaction \citep{1992ApJ...395..540C, 2016ApJ...821...36K}, in which the only relevant dynamical timescale is the pulsar spin-down time.  The main distinction between this problem and the pulsar-ejecta interaction is that the energy is now injected within a small opening angle.

Given this, the success or failure of the GRB jet cannot depend on the engine duration explicitly, only on the total injected energy, and presumably the opening angle within which the energy is injected.
We write the ability of the jet to break out of the dynamical ejecta  in terms of $E_{\rm crit}$, the minimum energy required for successful breakout:
\begin{equation}
E_j > E_{\rm crit} ~~{\rm (successful ~breakout).}
\end{equation}
If the total energy injected is below $E_{\rm crit}$, then all of the energy is adiabatically absorbed into the kinetic energy of the dynamical ejecta.  It is reasonable to expect that $E_{\rm crit}$ must be some fraction of the kinetic energy of the dynamical outflow.  Exactly what this fraction is should depend on $\theta_j$, the opening angle of the injected jet.  A likely scaling is
\begin{equation}
E_{\rm crit} \propto \theta_j^2 E_{\rm ej},
\end{equation}
based on the amount of energy subtended by the opening angle $\theta_j$.  Prior to performing numerical calculations, however, it is not obvious whether this scaling is correct, as the opening angle of the cocoon is not necessarily proportional to the opening angle of the injected jet.   Moreover, there is a second energy scale in the problem, $M_{\rm ej} c^2$.  For highly collimated jets  the speed of the head of a jet propagating through a medium is set by the momentum flux of the jet, while the speed of the cocoon spreading laterally is set by the energy content of the cocoon \citep{1989ApJ...345L..21B}.   In this case one can show that the jet propagation will depend on both $E_{\rm ej}$ and $M_{\rm ej} c^2$ independently.  For our problem, however, the mean ejecta velocities from neutron star mergers  live in a narrow range $\sim 0.1-0.3 c$, set by the escape velocity from a neutron star.   Thus $E_{\rm ej}$ and $M_{\rm ej} c^2$ are roughly proportional to each other.   For simplicity, we will thus present our results in terms of $E_{\rm ej}$, largely keeping the mean ejecta velocity $\sim (E_{\rm ej}/M_{\rm ej})^{1/2}$ fixed.   Additional tests using different mean ejecta velocities $\sim 0.1-0.4 c$ show that our conclusions are all essentially unchanged within this range (section \ref{sec:results}).



\subsection{Breakout Time}
\label{sec:analy-time}

\begin{figure*}
\epsscale{1.1}
\plotone{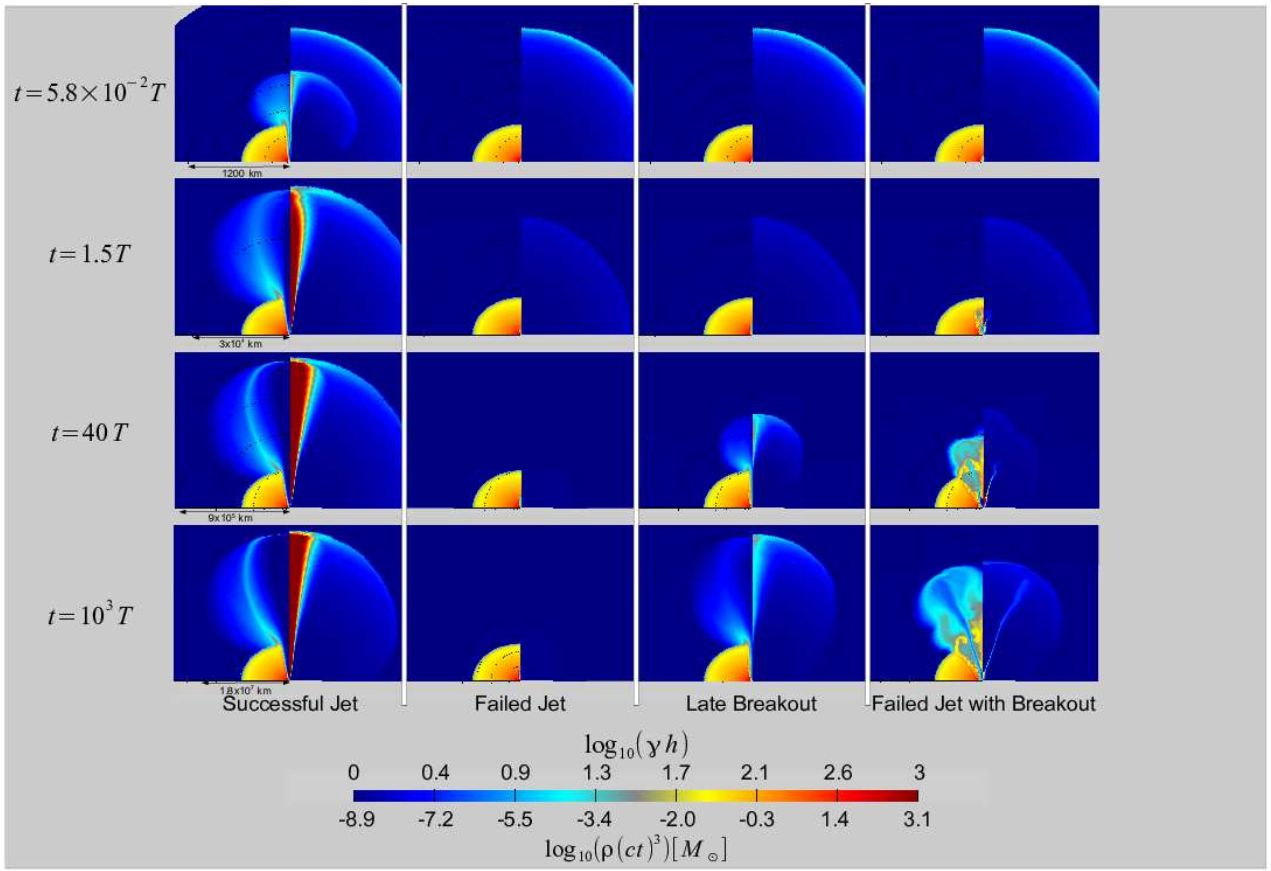}
\caption{Various scenarios for jet propagation in a homologous ejecta.  Each panel displays density on the left, and $\gamma h$ on the right.  The four columns represent different injected energies.  Each row represents a different time, as indicated on the left in units of the engine duration, $T$.  We take the ejecta energy to be $10^{51}$ ergs to present results in physical units. The leftmost column shows an early breakout, with an opening angle of $\theta_j = 0.1$ and a large injected energy of $E_j = 10^{50}$ ergs, much larger than the critical energy needed to breakout $E_{\rm crit} \approx 5 \times 10^{47}$ ergs (eq. \ref{eqn:ecrit}).  The second column injects a much lower energy, $4 \times 10^{47}$ erg (also with $\theta_j = 0.1$), below $E_{\rm crit}$.  This model initially drives narrow jet, but it is too weak and no wave or shock makes it out of the ejecta.  The third column shows a ``late breakout'', with $\theta_j = 0.1$ and $E_j = 1.6 \times 10^{48}$ ergs, just above $E_{\rm crit}$.  This results in a collimated and relativistic jet, with Lorentz factor $\sim 30$, which breaks out an order of magnitude in time after the engine has turned off.  The rightmost column has a high injected energy ($E_j = 10^{50}$ ergs) and a wide injection angle ($\theta_j = 0.4$).  The wide injection angle results in a ``choked jet'', but the very high energy results in a successful shock/cocoon breakout, even though the choked injection effectively becomes spherical.  The wide injection angle is necessary to choke such an energetic jet.
\label{fig:pretty}}
\end{figure*}

The breakout time can be estimated as in the pulsar problem \citep{1989ApJ...341..867C}, though with geometric corrections due to the jet opening angle.  We will proceed in a little bit less detail, to calculate the scalings but  without calculating the dimensionless coefficients in front, which would depend on the detailed structure of the dynamical ejecta.   Our numerical results will provide a calibration of these dimensionless coefficients.

The engine injects a jet with opening angle $\theta_j$, power $L$ and duration $T$, with total injected energy $E_j = L T$.  
By scale invariance, every length and timescale we calculate will be proportional to $T$.  Specifically, one should be able to write the breakout time as

\begin{equation}
t_{\rm bo} = T f( E_j ),
\end{equation}
for some dimensionless function $f$ of the injected energy.

One could envision a thought-experiment where the duration is essentially infinite, so that the engine is only specified by a constant power.  In this case, the total input energy is effectively infinite, so there is always a successful breakout.  After breakout, however, one is free to shut off the engine at any time, and therefore, for a given power, the minimal engine duration $T$ necessary for breakout is given by the relationship $L T = E_{\rm crit}$.
One can turn this formula around to estimate the breakout time; $t_{\rm bo} \sim E_{\rm crit}/L$ or

\begin{equation}
t_{\rm bo} \sim T E_{\rm crit}/E_j.
\end{equation}
As we shall show, this expression works for jets which break out before the engine has begun to turn off, so that the jet power can be considered a constant.

In this work, we also find a ``late breakout" regime in which the jet breaks out after the engine is turned off, $t_{\rm bo} > T$.  In this case, essentially all of $E_j$ has been delivered to the cocoon prior to breakout, which continues to collimate the jet. The jetted outflow has a characteristic terminal velocity $v_{\rm sh}$ which takes time to overtake the fastest-moving ejecta.  Since the breakout can occur well after the engine turns off, this regime is analogous to a collimated explosion in a homologous outflow.  We will calibrate the breakout time for this regime empirically in section \ref{sec:results}.












\subsection{Thermal Energy in the Cocoon}

Some fraction of the engine power will be converted into thermal energy of the cocoon as the jet propagates through the ejecta.  This, in addition to thermalized energy from radioactive decay, can potentially help power the thermal electromagnetic counterpart to neutron star mergers.  The engine likely outputs a primarily kinetic (or magnetic) flow which thermalizes in a shock when colliding with the ejecta.  Imagine that some fraction $\eta$ of the engine power is instantly thermalized in this shock.
The thermal energy as a function of time is then given by the equation

\begin{equation}
\frac{d E_{\rm Th}}{dt} = \eta L(t) - E_{\rm Th}/t,
\label{eqn:dEdt}
\end{equation}
where the second term is adiabatic conversion to kinetic energy in the homologous phase (a radiation-dominated equation of state is assumed).  Note that before the jet and cocoon themselves reach  homology the actual work done on the ejecta is not the homologous expression $-E_{\rm Th}/t$, but rather $p_c dV/dt$ (where $p_c$ and $dV/dt$ are the pressure and rate of change of volume of the cocoon); this in turn depends on how the jet head and cocoon propagate through the ejecta.   In equation \ref{eqn:dEdt} we are absorbing this physics into  
the factor $\eta$, which encapsulates the thermalization of the injected jet energy in the cocoon and the expansion of a cocoon driven shock into the ejecta.  We will show below that analytical considerations result in a value of $\eta \approx 0.8$, close to the value we find in our numerical calculations.

Equation \ref{eqn:dEdt} can be straightforwardly solved:
\begin{equation}
E_{\rm Th}(t) = \frac{\eta}{t} \int_0^t t' L(t') dt'.
\label{eqn:integral}
\end{equation}

If the jet breaks out, however, the forward shock detaches and the engine no longer efficiently thermalizes its energy.  To first approximation, the first term in (\ref{eqn:dEdt}) then vanishes and the thermal energy drops as $t^{-1}$:
\begin{equation}
E_{\rm Th}(t) = E_{\rm Th}(t_{\rm bo}) \frac{ t_{\rm bo} }{t} \ \ \ (t > t_{\rm bo}).
\end{equation}
In this study, we consider an engine model with a smooth cutoff at late times:
\begin{equation}
L(t) = \frac{E_j / T}{(1 + t/T)^2},
\label{eqn:en_power}
\end{equation}
so that equation \ref{eqn:integral} yields
\begin{equation} 
E_{\rm Th}(t) = \left\{ \begin{array}
				{l@{\quad \quad}l}
				\eta E_j \frac{T}{t} \left[ {\rm ln}( 1 + t/T ) - \frac{ t }{ T + t } \right]		& t < t_{\rm bo} \\  
    			\eta E_j \frac{T}{t} \left[ {\rm ln}( 1 + t_{\rm bo}/T ) - \frac{ t_{\rm bo} }{ T + t_{\rm bo} } \right]		& t > t_{\rm bo} \\ 
    			\end{array} \right.    
\label{eqn:therm1}
\end{equation}

If the shock breaks out at sufficiently early times, then one can ignore the details of how quickly the engine shuts off.  However, there is an additional complication, as the engine can still deposit thermal energy (at a diminished efficiency) onto the sides of the tunnel.  In this case, the first term in (\ref{eqn:dEdt}) does not strictly vanish, but is instead replaced with a diminished efficiency, $\eta_2$.  This would lead to the solution

\begin{equation} 
E_{\rm Th}(t) = \left\{ \begin{array}
				{l@{\quad \quad}l}
				\eta E_0(t)		& t < t_{\rm bo} \\  
    			\eta E_0(t_{\rm bo}) \frac{t_{\rm bo}}{t} + \eta_2 E_0(t) 	& t > t_{\rm bo} \\ 
    			\end{array} \right.   
    			\label{eqn:therm2} 
\end{equation}
where $E_0(t) = E_j \frac{T}{t} \left[ {\rm ln}( 1 + t/T ) - \frac{t}{T + t} \right]$.  The details of this formula are of course sensitive to our choice of $L(t)$, i.e. precisely how the power decays with time.  More generally, a good estimate for the late-time thermal energy is

\begin{equation} 
E_{\rm Th}(t) \approx \left\{ \begin{array}
				{l@{\quad \quad}l}
    			\eta_2 E_j \frac{T}{t} & {\rm early ~breakout} \\ 
				\eta E_j \frac{T}{t} & {\rm late ~breakout} \\  
    			\end{array} \right.  
    			\label{eqn:therm3}  
\end{equation}
In this study, we find the transition between early and late breakout occurs around $E_j \sim 30 E_{\rm crit}$ (where $E_{\rm crit}$ is the transition between failed and successful jets), and that $\eta = 0.8$, $\eta_2 \approx 0.02 \eta$ for our models.

\cite{1992ApJ...395..540C} calculate the fraction of thermalized energy for a spherically injected engine in a homologous outflow.

\begin{equation}
E_{\rm Th} = \frac{5-k}{11-2k} L t = \frac12 \eta L t
\end{equation}
for a pulsar wind injected into a density profile with $\rho \propto r^{-k}$ (their equation 2.8).  This result is weakly dependent on $k$, and leads to a thermalization efficiency of $\eta \approx 0.8$ for most of the propagation of the jet (using $k=2.5$).  Our numerical results are consistent with this value (see the analytical curves in Figure \ref{fig:therm}, which assume $\eta = 0.8$).

\section{Numerical Set-Up} \label{sec:numerics}

We numerically study the interaction of a jet with a homologous outflow using the equations of two-dimensional (2D) axisymmetric relativistic hydrodynamics:

\begin{equation} 
\partial_{\mu} ( \rho u^{\mu} ) = S_D \label{eqn:claw1}
\end{equation} 
\begin{equation} 
\partial_{\mu} ( \rho h u^{\mu} u^{\nu} + P g^{\mu \nu} ) = S^{\nu} \label{eqn:claw2}
\end{equation} 
where $\rho$ is proper density, $\rho h = \rho + \epsilon + P$ is enthalpy density, $P$ is pressure, $\epsilon$ is the internal energy density, $u^{\mu}$ is the
four-velocity, and $g^{\mu \nu}$ is the Minkowski metric (the equations are expressed in units such that $c=1$).  The source terms $S_D$ and $S^{\nu}$ will be used to model the injection of mass, momentum, and energy by the central engine on small scales.  The equation of state is assumed to be radiation dominated: $\epsilon = 3 P$. 

The equations are integrated using the moving-mesh hydrodynamics code, JET \citep{2011ApJS..197...15D, 2013ApJ...775...87D}.  The JET code's moving mesh makes it effectively Lagrangian, which allows for accurate resolution of supersonic and relativistic flows.  The inner and outer boundaries are also moved during the flow's evolution, so that the dynamic range covered at any given time is kept modest.  The spatial resolution is given by $\Delta \theta \sim \Delta r / r \sim 0.01$.  This value is not exact, as zones can expand and contract, but zones are refined or de-refined so as to maintain aspect ratios of order unity.  Additionally, zones are more concentrated near the symmetry axis, so that resolution there is about twice as high ($\Delta \theta \sim \Delta r / r \sim 0.005$).

Initial conditions are modeled after the numerical output of a neutron star merger simulation \citep{2013PhRvD..87b4001H, 2014ApJ...784L..28N}.  It is an outflow described by the following:

\begin{equation} 
\rho( r , \theta , 0 ) = \left\{ \begin{array}
				{l@{\quad \quad}l}
				f(\tilde r) + \rho_{\rm atm}(r) & \tilde r < R_0 	\\  
    			\rho_{\rm atm}(r) & \tilde r > R_0 	\\ 
    			\end{array} \right.    
\end{equation}
\begin{equation} 
v_r( r , \theta , 0 ) = \left\{ \begin{array}
				{l@{\quad \quad}l}
				r/t_0 & \tilde r < R_0 	\\  
    			0 & \tilde r > R_0 	\\ 
    			\end{array} \right.    
\end{equation}
\begin{equation}
f(r) = \rho_c { (1-r/R_0)^{3/4} \over 1 + (r/R_1)^{2.5} } 
\end{equation}
\begin{equation}
\rho_{\rm atm}(r) = \rho_{\rm trans} e^{-r/R_0} + \rho_{\rm ISM}
\end{equation}
\begin{equation}
\tilde r^2 = r^2 ( w^{4/3} ~{\rm cos}^2 \theta + w^{-2/3} ~{\rm sin}^2 \theta )
\end{equation}

\noindent
where $\rho_c = 97 M_0 / R_0^3$, $R_1 = 0.06 R_0$, $\rho_{\rm trans} = 10^{-8} M_0/R_0^3$, $\rho_{\rm ISM} = 10^{-20} M_0/R_0^3$, and $R_0 = v_{\rm max} t_0$, with $v_{\rm max} = 0.3 c$.  The engine is initiated at a very early time $t_0 = 10^{-3} T$, where $T$ is the engine duration.  The ejecta model above corresponds to a mass-weighted velocity of the ejecta of $0.1 c$ and kinetic energy-weighted velocity of the ejecta of $0.18 c$.

In interpreting our results, the mass $M_0$ of the cloud is assumed to have a fiducial value of $0.07 M_{\sun}$ which gives a kinetic energy of $10^{51}$ erg. The dynamics of the interaction are, however, scale-invariant, so one could equally well assume, say, $M_0 = 0.01 M_{\sun}$ and $E_{\rm ej} = 1.4 \times 10^{50}$ erg.  In cgs units, the external density $\rho_{\rm ISM} \approx 5 \times 10^{-17}$ g/cm$^3$ is much larger than typical ISM densities of $\sim 1$ proton per cm$^3$, but this is irrelevant for the present study, as we are only concerned with the interaction between the jet and the ejecta, so this density should be considered an artificial atmosphere, with low enough density not to affect the dynamics.  The parameter $w = 1.3$ is the aspect ratio of the oblate cloud.  The choice $w = 1.3$ has an impact on whether or not the jet can break out of the ejecta \citep{2015ApJ...813...64D}, and therefore this affects any dimensionless coefficients reported in this study, but we expect that the overall scalings are independent of the details of the ejecta model.

\begin{figure}
\epsscale{1.2}
\plotone{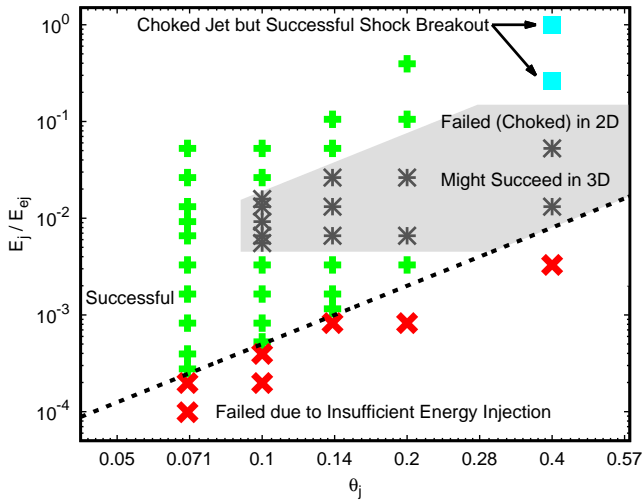}
\caption{Diagram illustrating whether or not the jet in a given model succeeds at breaking out of the ejecta, as a function of the injection opening angle $\theta_j$ and the injected energy $E_j$.  Green points indicate successful breakouts.  Grey points are jets which were energetic but exhibited a significant cork along the symmetry axis that the jet collided with, causing failure in 2D (these jets might instead be successful in 3D; see text for details).  Red x's indicate jets that failed due to insufficient energy injection.   Cyan squares were  cases which exhibited shock/coccon breakout even with a choked jet; these require both high energy and a wide injection angle in order to choke such an energetic jet.  $E_{\rm crit}(\theta)$ (eq. \ref{eqn:ecrit}) delineating success from failure is indicated with the dashed curve. 
\label{fig:failure} }
\end{figure}

The engine power is prescribed by equation (\ref{eqn:en_power}), $L(t) = (E_j/T)/(1+t/T)^2$, which is a constant power for $t \ll T$, but decays as $E_j T / t^2$ at late times.  The power-law tail is non-standard in most numerical calculations of GRB jets; typically jet power is imposed for some given duration with an instantaneous shut-off.  Nature is more likely to be well-modeled as a power-law decay than to have a sharp cutoff.  The exponent of $-2$ is largely arbitrary, but it could mimic late-time power input by a central engine, e.g., magnetar spin-down or fallback accretion.

\section{Results} \label{sec:results}

The two dimensionless variables which are varied in this study are $E_j / E_{\rm ej}$ (the ratio of injected energy to the kinetic energy of the outflow) and $\theta_j$, the opening angle of the injection (which may be different from the opening angle of the final jet).  We have also performed additional runs varying $v_{\rm max}$ in the ejecta model (\S \ref{sec:numerics}).  This effectively varies the energy scale $E_{\rm ej}$ while keeping the mass fixed and confirms that the scaling with $E_{\rm ej}$ at fixed $M_{\rm ej} c^2$ (or vice-versa) is weak over the range of   
mean ejecta velocities $\sim 0.1 - 0.4$ c expected in neutron star mergers.   

Figure \ref{fig:pretty} shows the four classes of solutions that we find depending on the values of the key dimensionless numbers $E_j / E_{\rm ej}$ and $\theta_j$.   These are, from left to right:  successful  jet breakout from the homologous ejecta on a timescale $\lesssim T$, the engine duration, failed jets, late breakout on timescales $\gtrsim T$, and failed jets in which the forward shock and cocoon driven into the ejecta by the jet nonetheless breaks out.   In what follows we delineate the regimes in which these different solutions occur and some of their key properties.   

\subsection{Success vs. Failure}

We argued in \S \ref{sec:anly} that the ability of a jet to break out of the ejecta should depend on the jet energy relative to a critical energy $E_{\rm crit}$, with
\begin{equation}
E_{\rm crit} = \kappa E_{\rm ej} \theta_j^2.
\label{eqn:ecrit}
\end{equation}
where $\kappa$ is some dimensionless constant.  Figure \ref{fig:failure} shows success vs. failure for a wide range of models with variable energy and opening angle.  A successful jet is defined here as one which emerges from the outflow collimated and relativistic.  A successful shock breakout is defined to occur if any shock at all makes it to the edge of the ejecta.  For all except the cyan points in Figure \ref{fig:failure}, successful shock breakout necessitated a successful jet.  We find the $\theta_j^2$ scaling to be accurate for determining success of the jet, with the coefficient $\kappa = 0.05$ in equation (\ref{eqn:ecrit}).  The value of this coefficient should depend on the details of the initial density profile and the engine power as a function of time.  

One important caveat of these numerical calculations is that they have been performed in 2D, and some of these jets fail due to ``corking" \citep[the jet is choked because a massive cork or plug is trapped ahead of the jet, which deflects the jet away from the symmetry axis; this might not occur in 3D, see][]{2015ApJ...806..205D}.  These are labeled as such in Figure \ref{fig:failure}. Fortunately, a scaling for $E_{\rm crit}(\theta_j)$ can be discerned despite these possibly artifically choked jets.  

\subsection{Breakout Time}

\begin{figure}
\epsscale{1.2}
\plotone{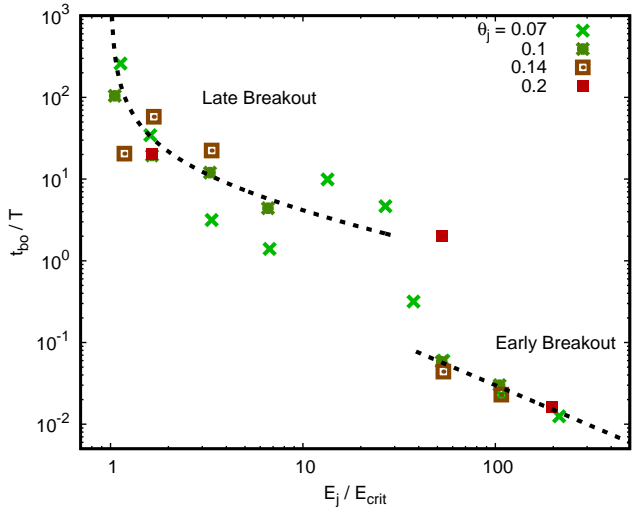}
\caption{Jet breakout time for all successful models, compared with the prediction given by equation (\ref{eqn:tbo}).  Jets with $E_j > 30 E_{\rm crit}$ exhibit an ``early breakout", where the engine remains on after breakout.  Jets with $E_j \lesssim 30 E_{\rm crit}$ exhibit ``late breakout," breaking out well after the engine has ceased.  Late breakout of the jet can produce a delay between gravitational wave and gamma-ray emission like that seen in GW170817. 
\label{fig:times} }
\end{figure}

Breakout times of successful jets are shown in Figure \ref{fig:times}.  Breakout times are consistent with the following expression:

\begin{equation} 
t_{\rm bo} = \left\{ \begin{array}
				{l@{\quad \quad}l}
				\alpha_1 T \frac{E_{\rm crit}}{E_j}		& {\rm early ~breakout}	\\  
    			\alpha_2 \frac{ T }{ \sqrt{E_j/E_{\rm crit}} - 1 }	& {\rm late ~breakout} \\ 
    			\end{array} \right.
    			\label{eqn:tbo}
\end{equation}
where the ``early breakout" scaling was estimated in 
\S \ref{sec:analy-time}, but the ``late breakout" scaling is entirely empirical.  For $E_j \gtrsim 30 E_{\rm crit}$, breakout times follow the ``early breakout" scaling  with a coefficient $\alpha_1 \simeq 3$ while for $E_{\rm crit} < E_j \lesssim 30 E_{\rm crit}$, there is more scatter, but breakout times reasonably track the ``late breakout" scaling, with $\alpha_2 \simeq 9$. 

\begin{figure}
\epsscale{1.2}
\plotone{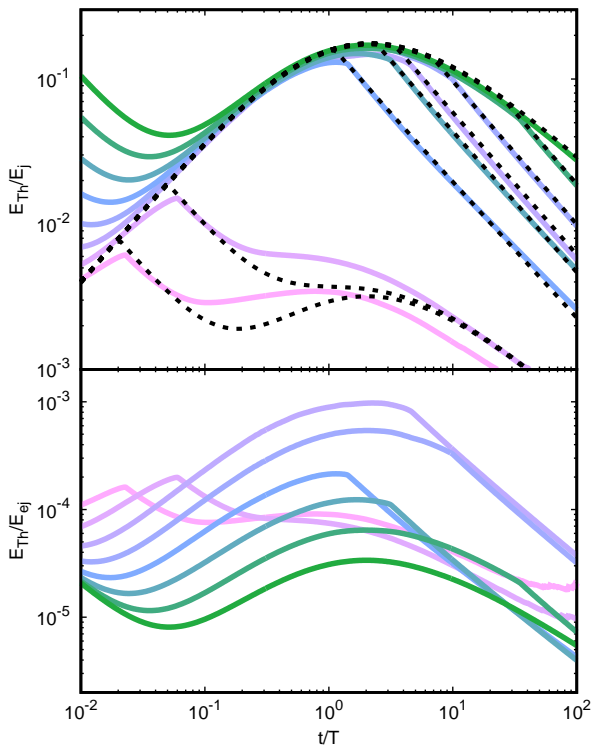}
\caption{Total thermal energy as a function of time, for models with $\theta_j = 0.07$, and with jet energies ranging from $8 \times 10^{47}$ ergs (green) to $5 \times 10^{49}$ ergs (pink), assuming an ejecta kinetic energy of $10^{51}$ ergs.  No radioactive heating is accounted for.  Failed jets precisely follow the analytic expression in equation (\ref{eqn:therm2}) with thermalization efficiency $\eta = 0.8$ (dashed curves).  Successful jets depart from this scaling, as the forward shock breaks out of the ejecta, and the thermalization is suppressed after this point, leading to a sudden transition to a $t^{-1}$ scaling at $t_{\rm bo}$.  Jets that break out early have a second, much less efficient injection of thermal energy, due to the engine heating the walls of the tunnel.
\label{fig:therm} }
\end{figure}

\begin{figure}
\epsscale{1.2}
\plotone{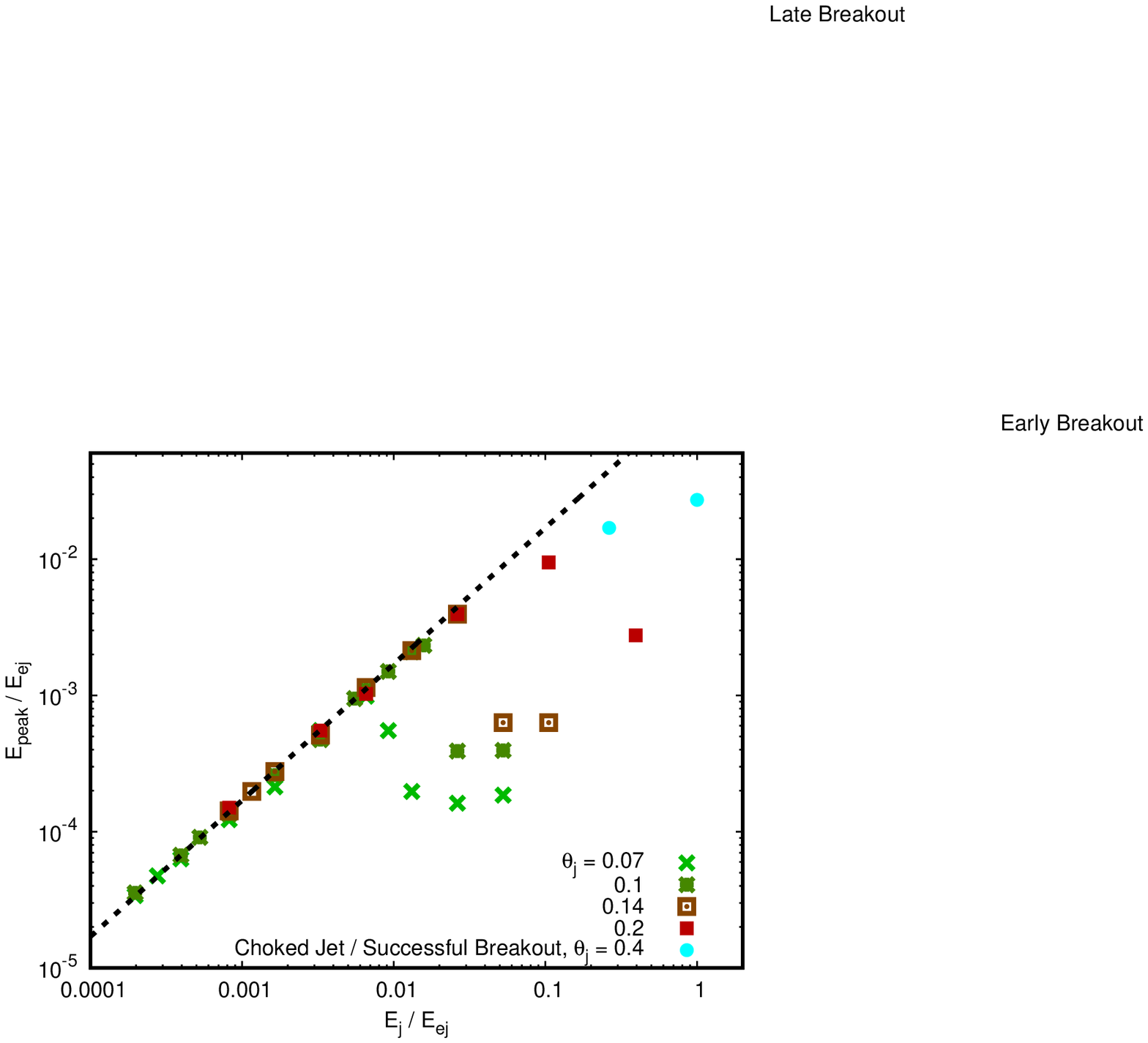}
\caption{Peak thermal energy due to jet heating for the various models considered.  Figure \ref{fig:therm} shows that this peak thermal energy is achieved on a timescale comparable to the engine duration. Failed jets and late breakouts cluster along the curve $E_{\rm peak} \simeq 0.17 E_j$.  Early breakouts result in a reduced peak thermal energy.  The most dramatic jet models had peak thermal energy of a few times $10^{-2} E_{ej} \sim 10^{49}$ ergs (taking $E_{ej} = 10^{51}$ ergs).   This is significantly less than the thermal energy supplied by radioactive heating of the ejecta. 
\label{fig:peak} }
\end{figure}

\subsection{Thermal Energy}

The total thermal energy in the computational domain is shown as a function of time in Figure \ref{fig:therm}.  In the upper panel, this is compared with the analytical form given by equation (\ref{eqn:therm3}).  For failed jets (green curves), the thermal energy follows the predicted formula precisely, with a thermalization efficiency $\eta = 0.8$.  Successful jets with early breakout (pink curves) have a sudden transition at breakout, where the thermal energy drops and thermalization is no longer efficient because the jet  vents through the channel carved in the ejecta.  Jets with an early breakout exhibit a second increase in thermal energy due to the thermalization of the jet onto the sides of the tunnel.  This comes with a reduced efficiency $\eta_2 \approx 0.02 \eta$, though this efficiency could be higher if the jets precess or the jet opening angle varies in time due to variability in the  accretion disk driving the jet.   After the engine shuts off, i.e., for $t \gtrsim T$, the thermal energy evolution transitions
to a $t^{-1}$ decline consistent with adiabatic conversion from thermal to kinetic energy.


The lower panel in Figure \ref{fig:therm} shows the thermal energy as a function of time in units of the ejecta kinetic energy rather than the input jet energy.   This highlights that it is  difficult to put a significant amount of thermal energy into the cocoon/ejecta via a jet even when the total energy of the jet approaches the initial kinetic energy of the ejecta (redder curves).  This is primarily because high jet energies result in an early breakout, cutting off the thermalization and leading to most of the jet energy escaping the ejecta.   

Figure \ref{fig:peak} shows the peak thermal energy in each of our models.  This peak is maximized for failed jets and late breakouts with large opening angles; for such cases, the peak thermal energy is $E_{\rm peak} \approx 0.17 E_j$, where the factor $0.17$ can be determined from the peak of equation (\ref{eqn:therm1}).  The exact value of this coefficient at the factor of few level will depend on the precise form of $L(t)$ and the ejecta structure.  Figure \ref{fig:peak} shows that only for very large opening angles and jet energies approaching $E_{ej} \sim 10^{51}$ ergs can the jet deposit an appreciable thermal energy in the ejecta.  More typically, $E_{\rm peak} \ll E_{\rm ej}$.

\section{Discussion} \label{sec:disc}

The energetics of short GRBs are less constrained than those of long-duration bursts, because the opening angles of short bursts are usually more uncertain.  However,  current estimates from afterglow observations are that  typical energies are in the range $10^{49} - 10^{50}$ erg, with opening angles of $\theta_j \sim 0.05 - 0.4$ \citep{2015ApJ...815..102F}.  An independent (but model dependent) constraint on the opening angle is provided by the relative rate of short GRBs and neutron star mergers, which favors beaming factors of $\sim 100$ and thus $\theta_j \sim 0.1$.  

Note that opening angles measured from afterglow observations utilize the ``jet break" and  may be different from the  opening angle injected by the central engine, which is how $\theta_j$ is defined in the present study. Similarly, the duration of the burst might not be the same as the duration of the engine $T$, though we assume that a typical value of $T$ should be in the typical observed duration range of $0.1 - 1$ second.  All of the calculations in this work assumed that the engine is initiated at  early times, $\ll T$ and $\ll t_{\rm bo}$.  This assumption is reasonable, as an accretion disk should form on millisecond timescales. Even if a hypermassive neutron star survives for $\sim 0.1-1$ sec before collapsing to a black hole, it is unlikely that much mass will remain at later times to accrete onto the black hole given the short viscous times in the disk around the hypermassive neutron star (\citealt{2014MNRAS.441.3444M} demonstrate this explicitly in idealized axisymmetric hydrodynamic simulations).

For jet energies and opening angles at the upper limits suggested by observations of short GRBs ($E_{\rm j} \sim 10^{50}$ ergs $\sim 0.1 \, E_{\rm ej}$ and $\theta_j \sim 0.2-0.4$), Figures \ref{fig:therm} and \ref{fig:peak} show that the peak thermal energy due to the jet propagating through the ejecta is $\sim 10^{-3}-10^{-2} E_{\rm ej} \sim 10^{48}-10^{49}$ ergs (where the latter numbers assume our fiducial $E_{\rm ej} = 10^{51}$ ergs).  Moreover, this energy is reached on a timescale comparable to the engine duration $T \sim 0.1-1$ sec.  For comparison, the r-process produces a thermal energy of $\simeq 1-3$ MeV per nucleon on a timescale of about 1 sec \citep{2010MNRAS.402.2771M}, with the heating then decaying as a power-law at later times.  For our fiducial ejecta mass of $0.07 M_\odot$ this corresponds to a thermal energy from the r-process of $1-3 \times 10^{50} {\rm ergs} \gtrsim 0.1 E_{\rm ej}$ on a timescale of 1 sec.\footnote{This conclusion holds even if $E_{\rm ej}$ is lower, say $E_{ej} \sim 10^{50}$ ergs and $M_{ej} \sim 0.01 M_\odot$; in this case $E_{\rm j} \sim 10^{50} \, {\rm ergs} \, \sim E_{\rm ej}$ leads to a maximum thermal energy of $\sim 10^{47}-10^{48}$ ergs for $\theta_j \sim 0.2-0.4$ (Fig \ref{fig:peak}), while r-process heating produces an energy in the first few seconds of $\sim 3 \times 10^{49}$ ergs.}  Figure \ref{fig:peak} thus shows that r-process heating of the ejecta in neutron star mergers on second timescales dominates over jet heating in the entire parameter space of numerical models studied in this paper (usually by a large factor).   R-process heating will thus set the overall luminosity of any associated thermally powered transient.   Note that the dominance of r-process heating over jet heating is usually discussed in the context of day-timescale kilonova emission.   Here we have shown that this is true in most cases even on {\em second} timescales.   We do find, however, that for a small amount of mass, the energy per unit mass from shock heating exceeds that produced by radioactive heating in the first few seconds.   The implications of this for kilonova lightcurves will be explored in future work.


For almost all of our parameter choices, we find that shock breakout of the cocoon only occurs if the jet itself is able to escape the ejecta.  This is in contrast with some of the shock breakout scenarios presented for GRB 170817A \citep[e.g.][]{2018MNRAS.tmp.1404G} which assume cocoon breakout after a choked/failed jet.  The regime of shock breakout with a choked jet is only realized for large jet energies and large jet opening angles (see the right panel of Fig. \ref{fig:pretty} and the cyan points in Fig. \ref{fig:failure}), which are indeed the parameters used in \citet{2018MNRAS.tmp.1404G}.  By contrast, for lower energies and opening angles, which appear to be somewhat more typical for short GRBs, the jet and cocoon both likely break out (Fig. \ref{fig:failure}).   


We find that the timescale for the jet to breakout of the ejecta scales inversely with the jet power for $E_j \gg E_{\rm crit}$.  For energies just above $E_{\rm crit}$, however, the difference between the shock velocity due to the jet and the homologous ejecta velocity is small, resulting in a late breakout on timescales that can be significantly longer than the engine duration $T$.  
In particular, reasonable engine parameters consistent with short-duration GRBs can be consistent with the 1.7 second delay between gravitational waves and gamma rays in GW 170817.  For example, a delay of $\sim 3-10 \, T$ occurs  in our late breakout regime with $E_j \sim 10 E_{\rm crit}$ (Fig. \ref{fig:times}).  Taking $\theta_j \sim 0.1$, this corresponds to $E_j \sim 10^{-2} E_{\rm ej} \sim 10^{49}$ ergs, well within reasonable short GRB parameters.  The hypothesis that the delay between gamma-rays and gravitational waves in GW 170817 is caused by the long breakout time can be tested by future observations. For a reasonable spread in engine parameters, it is likely that future EM counterparts to neutron star mergers will be separated into two classes: bursts with a substantial delay in gamma-rays and those without.  The bursts without delay would have larger ratios of jet energy to ejecta energy.  Mergers producing very little dynamical ejecta and thus fainter kilonovae would have very little delay.

The calculations presented in this paper are scale-free and can be re-scaled to different assumed dynamical ejecta and jet properties.   We have used a kinetic energy for the quasi-spherical dynamical ejecta of $E_{ej} = 10^{51}$ ergs to anchor our discussion.   This is consistent with observations of the optical-IR counterpart to GW170817 (e.g., \citealt{2017ApJ...851L..21V}).   Quasi-spherical dynamical ejecta with this energy can be produced in binary neutron star mergers for sufficiently small neutron star radii (e.g., \citealt{2016MNRAS.460.3255R}).    If, however, the quasi-spherical ejecta energy is lower -- which is  consistent with GW170817 observations if the blue kilonova component is produced by, e.g., accretion disk or magnetar winds -- a jet of a given energy and opening angle will break out more easily and will deposit less thermal energy into the cocoon as a result (Figs. \ref{fig:times} \& \ref{fig:peak}).

By using short GRB observations to motivate engine parameters for gravitational wave selected events, we are assuming that at least some short GRBs are produced by binary neutron star mergers.  We believe that this empirical calibration is reasonable even if short GRBs are instead due to neutron star-black hole mergers, because the mass and angular momentum content of the accretion disks are not that different.   By contrast, if short GRBs are produced by an entirely different source (e.g., accretion-induced collapse of white dwarfs to neutron stars), then calibrating engine parameters of gravitational wave detected events to short GRB observations may be inappropriate.  It is of course also possible that binary neutron star mergers do produce most short GRBs, but that there is a large dispersion in short GRB engine properties and that GW 170817 just happened to be somewhat unusual relative to the bulk of the population (e.g., some short GRBs do not show any evidence for a jet break, consistent with a large opening angle of $\theta_j \sim 0.4$; see, e.g., GRB 050724A in \citealt{2015ApJ...815..102F}.) Multi-messenger observations in the coming decade will likely empirically settle these questions.

\acknowledgments

We thank Brian Metzger, Juna Kollmeier and Tony Piro for stimulating conversations that helped to motivate much of this work.  We thank Wen-fai Fong, Ore Gottlieb, Andrew MacFadyen, and Rosalba Perna for useful conversations.   This work was supported, in part,  by the Theoretical Astrophysics Center at UC Berkeley.  This research was also funded by the Gordon and Betty Moore Foundation through Grant GBMF5076.  EQ was supported in part by a Simons Investigator award from the Simons Foundation.  HK is supported by a DOE Computational Science Graduate Fellowship under grant number DE-FG02-97ER25308.  Numerical computations utilized the Savio computational cluster resource provided by the Berkeley Research Computing program at the University of California, Berkeley (supported by the UC Berkeley Chancellor, Vice Chancellor of Research, and Office of the CIO).


\bibliographystyle{apj} 

\end{document}